\documentclass[showpacs,amsmath,amssymb,preprint]{revtex4}
\usepackage{epsfig,dcolumn,bm}

\begin{document}

\title{$\Delta K$, $\Lambda K$, and $\Sigma K$ states in the extended chiral SU(3) quark model}

\author{F. Huang$^{1,2,3}$}
\author{Z.Y. Zhang$^2$}
\affiliation{\small
$^1$CCAST (World Laboratory), P.O. Box 8730, Beijing 100080, PR China \\
$^2$Institute of High Energy Physics, P.O. Box 918-4, Beijing 100049, PR China\footnote{Mailing address.} \\
$^3$Graduate School of the Chinese Academy of Sciences, Beijing,
PR China}

\begin{abstract}
By use of the resonating group method, the $\Delta K$, $\Lambda
K$, and $\Sigma K$ states are further dynamically studied in the
extended chiral SU(3) quark model based on our previous work.
Similar to the results given by the original chiral SU(3) quark
model, the calculated results here still show that the
interactions of $\Delta K$ with isospin $I=1$ and $\Sigma K$ with
isospin $I=1/2$ are attractive, which can consequently lead to
$\Delta K$ and $\Sigma K$ quasibound states. When the channel
coupling of $\Lambda K$ and $\Sigma K$ is considered, the
calculated phase shifts show a sharp resonance between the
thresholds of these two channels with spin-parity $J^P=1/2^-$.
\end{abstract}


\maketitle

As we know, the non-perturbative QCD (NPQCD) effect is very
important in the light quark system. Since it is difficult to
seriously solve the NPQCD effect, QCD-inspired models are still
needed to connect the theoretical results and the experimental
observables. Among these models, the chiral SU(3) quark model has
been quite successful in reproducing the energies of the baryon
ground states, the binding energy of deuteron, the nucleon-nucleon
($NN$) scattering phase shifts, and the nucleon-hyperon ($NY$)
cross sections. Recently, we extended the chiral SU(3) quark model
to study the baryon-meson systems by solving a resonating group
method (RGM) equation. In Refs. \cite{fhuang04kn,fhuang04nkdk}, we
studied the kaon-nucleon ($KN$) scattering phase shifts, and a
satisfactory agreement with the experiment is obtained. Further,
in Refs. \cite{fhuang04nkdk,fhuang05lksk}, we dynamically studied
the structures of the $\Delta K$, $\Lambda K$, and $\Sigma K$
states. Our results show that the $\Delta K$ with isospin $I=1$
and the $\Sigma K$ with isospin $I=1/2$ have quite strong
attractions, which can consequently lead to $\Delta K$ and $\Sigma
K$ quasi-bound states with binding energy of about 2 and 17 MeV,
respectively. When the channel coupling of $\Lambda K$ and $\Sigma
K$ is considered, the calculated phase shifts show a sharp
resonance between the thresholds of $\Lambda K$ and $\Sigma K$
with spin-parity $J^P=1/2^-$ and width $\Gamma\approx 5$ MeV. The
strong attraction of $\Sigma K$ and the sizeable off-diagonal
matrix elements of $\Lambda K$ and $\Sigma K$ are responsible for
the appearance of this resonance. Our further analysis reveal that
the strong attractions of both $\Delta K$ with $I=1$ and $\Sigma
K$ with $I=1/2$ dominantly come from the $\sigma$ exchange and
color-magnetic force of the one-gluon exchange (OGE), and the
considerably large transition interaction from $\Lambda K$ to
$\Sigma K$ are dominantly offered by the OGE. In other words, the
OGE plays an important role in the $\Delta K$, $\Lambda K$, and
$\Sigma K$ systems in the chiral SU(3) quark model study.

For low-energy hadron physics, it remains a controversial problem
whether the gluon or the Goldstone boson is the proper effective
degree of freedom besides the constituent quark. Glozman and Riska
proposed that the Goldstone boson is the only other proper
effective degree of freedom \cite{glozman96}. But Isgur gave a
critique of the boson exchange model and insisted that the OGE
governs the baryon structure \cite{isgur02}. Nonetheless, it is
still a challenging problem in low-energy hadron physics whether
OGE or vector-meson exchange is the right mechanism or both of
them are important for describing the short-range quark-quark
interaction.

In Refs. \cite{lrdai03,fhuang05kne}, the chiral SU(3) quark model
is extended to include the coupling between the quark and vector
chiral fields. The OGE that dominantly governs the short-range
quark-quark interaction in the original chiral SU(3) quark model
is now nearly replaced by the vector-meson exchange. By use of
this model, we have obtained a satisfactory description of the
$NN$ and $KN$ scattering phase shifts.

The purpose of this work is to perform a further dynamical study
on the $\Delta K$, $\Lambda K$, and $\Sigma K$ systems in the
extended chiral SU(3) quark model based on Refs.
\cite{fhuang04nkdk,fhuang05lksk}. Let's first briefly review the
model (the detailed formula can be found in Ref.
\cite{fhuang05kne}). The total Hamiltonian of baryon-meson systems
can be written as
\begin{equation}
H=\sum_{i=1}^{5}T_{i}-T_{G}+\sum_{i<j=1}^{4}V_{ij}+\sum_{i=1}^{4}V_{i\bar
5},
\end{equation}
where $T_G$ is the kinetic energy operator for the c.m. motion,
and $V_{ij}$ and $V_{i\bar 5}$ represent the quark-quark and
quark-antiquark interactions, respectively,
\begin{equation}
V_{ij}= V^{OGE}_{ij} + V^{conf}_{ij} + V^{ch}_{ij},
\end{equation}
where $V_{ij}^{OGE}$ is the OGE interaction, $V_{ij}^{conf}$ is
the confinement potential, and $V^{ch}_{ij}$ is the chiral fields
induced effective quark-quark potential,
\begin{eqnarray}
V^{ch}_{ij} = \sum_{a=0}^8 V_{\sigma_a}({\bm r}_{ij})+\sum_{a=0}^8
V_{\pi_a}({\bm r}_{ij})+\sum_{a=0}^8 V_{\rho_a}({\bm r}_{ij}).
\end{eqnarray}
Here $\sigma_{0},...,\sigma_{8}$ are the scalar nonet fields,
$\pi_{0},..,\pi_{8}$ are the pseudoscalar nonet fields, and
$\rho_{0},..,\rho_{8}$ are the vector nonet fields. The
expressions of all the interactions can be found in the literature
\cite{fhuang04kn,fhuang04nkdk,fhuang05lksk,fhuang05kne}.

$V_{i \bar 5}$ in Eq. (1) includes two parts: direct interaction
and annihilation parts:
\begin{equation}
V_{i\bar 5}=V^{dir}_{i\bar 5}+V^{ann}_{i\bar 5},
\end{equation}
with
\begin{equation}
V_{i\bar 5}^{dir}=V_{i\bar 5}^{conf}+V_{i\bar 5}^{OGE}+V_{i\bar
5}^{ch},
\end{equation}
and
\begin{eqnarray}
V_{i\bar{5}}^{ch}=\sum_{j}(-1)^{G_j}V_{i5}^{ch,j}.
\end{eqnarray}
Here $(-1)^{G_j}$ represents the G parity of the $j$th meson. The
$q\bar q$ annihilation interactions, $V_{i\bar 5}^{ann}$, are not
included in this work because they are assumed not to contribute
significantly to a molecular state or a scattering process, which
is the subject of our present study.

All the model parameters are fixed before the calculation by some
special constraints, such as the mass splits between $N$, $\Delta$
and $\Lambda$, $\Sigma$, the stability conditions of $N$,
$\Lambda$ and $\Xi$, and the masses of $N$, $\Sigma$, and
$\overline{\Xi+\Omega}$. (For details See Refs.
\cite{fhuang04kn,fhuang04nkdk,fhuang05lksk,fhuang05kne}.) Their
values are listed in Table I, where the first set is for the
original chiral SU(3) quark model, the second and third sets are
for the extended chiral SU(3) quark model by taking
$f_{chv}/g_{chv}$ as $0$ and $2/3$, respectively. Here $g_{chv}$
and $f_{chv}$ are the coupling constants for vector coupling and
tensor coupling of the vector meson fields, respectively. $g_u$
and $g_s$ are the OGE coupling constants and $a^c$ represents the
strength of the confinement potential. All these three sets of
parameters can give a satisfactory description of the masses of
the baryon ground states, the binding energy of the deuteron, and
the $NN$ scattering phase shifts.

{\small
\begin{table}[htb]
\caption{\label{para} Model parameters. The meson masses and the
cutoff masses: $m_{\sigma'}=980$ MeV, $m_{\kappa}=980$ MeV,
$m_{\epsilon}=980$ MeV, $m_{\pi}=138$ MeV, $m_K=495$ MeV,
$m_{\eta}=549$ MeV, $m_{\eta'}=957$ MeV, $m_{\rho}=770$ MeV,
$m_{K^*}=892$ MeV, $m_{\omega}=782$ MeV, $m_{\phi}=1020$ MeV, and
$\Lambda=1100$ MeV.}
\begin{center}
\begin{tabular*}{85mm}{@{\extracolsep\fill}cccc}
\hline\hline
  & $\chi$-SU(3) QM & \multicolumn{2}{c}{Ex. $\chi$-SU(3) QM}  \\
  &   I   &    II    &    III \\  \cline{3-4}
  &  & $f_{chv}=0$ & $f_{chv}=2/3g_{chv}$ \\
\hline
 $b_u$ (fm)  & 0.5 & 0.45 & 0.45 \\
 $m_u$ (MeV) & 313 & 313 & 313 \\
 $m_s$ (MeV) & 470 & 470 & 470 \\
 $g_u^2$     & 0.781 & 0.067 & 0.143 \\
 $g_s^2$     & 0.865 & 0.212 & 0.264 \\
 $g_{ch}$    & 2.621 & 2.621 & 2.621  \\
 $g_{chv}$   &       & 2.351 & 1.973  \\
 $m_\sigma$ (MeV) & 595 & 535 & 547 \\
 $a^c_{uu}$ (MeV/fm$^2$) & 46.6 & 44.5 & 39.1 \\
 $a^c_{us}$ (MeV/fm$^2$) & 58.7 & 79.6 & 69.2 \\
 $a^c_{ss}$ (MeV/fm$^2$) & 99.2 & 163.7 & 142.5 \\
 $a^{c0}_{uu}$ (MeV)  & $-$42.4 & $-$72.3 & $-$62.9 \\
 $a^{c0}_{us}$ (MeV)  & $-$36.2 & $-$87.6 & $-$74.6 \\
 $a^{c0}_{ss}$ (MeV)  & $-$33.8 & $-$108.0 & $-$91.0 \\
\hline\hline
\end{tabular*}
\end{center}
\end{table}}

From Table I one can see that for both set II and set III, $g_u^2$
and $g_s^2$ are much smaller than the values of set I. This means
that in the extended chiral SU(3) quark model, the coupling
constants of OGE are greatly reduced when the coupling of quarks
and vector-meson field is considered. Thus the OGE that plays an
important role of the quark-quark short-range interaction in the
original chiral SU(3) quark model is now nearly replaced by the
vector-meson exchange. In other words, the mechanisms of the
quark-quark short-range interactions in these two models are quite
different.

With all parameters determined in the extended chiral SU(3) quark
model, the $\Delta K$, $\Lambda K$, and $\Sigma K$ states can be
dynamically studied in the framework of the RGM, a
well-established method for studying the interaction between two
composite particles.

The $\Delta K$ state has already been studied in Ref.
\cite{ssa04}, where the authors claimed that they find an
attractive interaction in the $\Delta K$ channel with $L=0$ and
$I=1$. This state has also been investigated in Ref. \cite{eek04}
based on the $\chi$-BS(3) approach. In Ref. \cite{fhuang04nkdk},
we study the $\Delta K$ state in the chiral SU(3) quark model and
find that the interaction of $\Delta K$ with isospin $I=1$ is
attractive. Such an attraction can make for a $\Delta K$
quasi-bound state with about 2 MeV binding energy. Our further
analysis shows that the attraction dominantly comes from the
$\sigma$ exchange and the color-magnetic force of OGE.

\begin{figure}[htb]
\vglue 2.3cm \epsfig{file=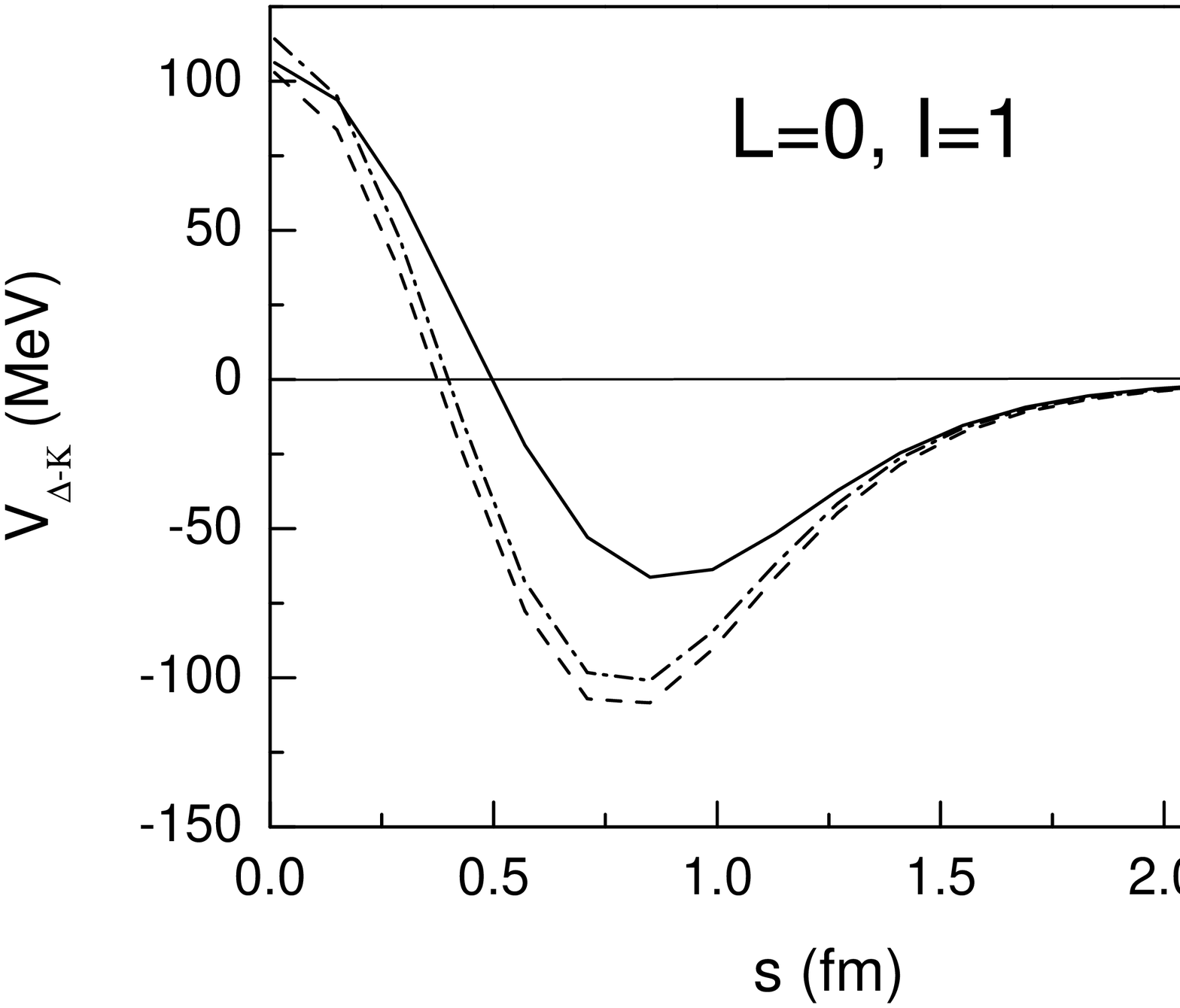,width=7.5cm}
\epsfig{file=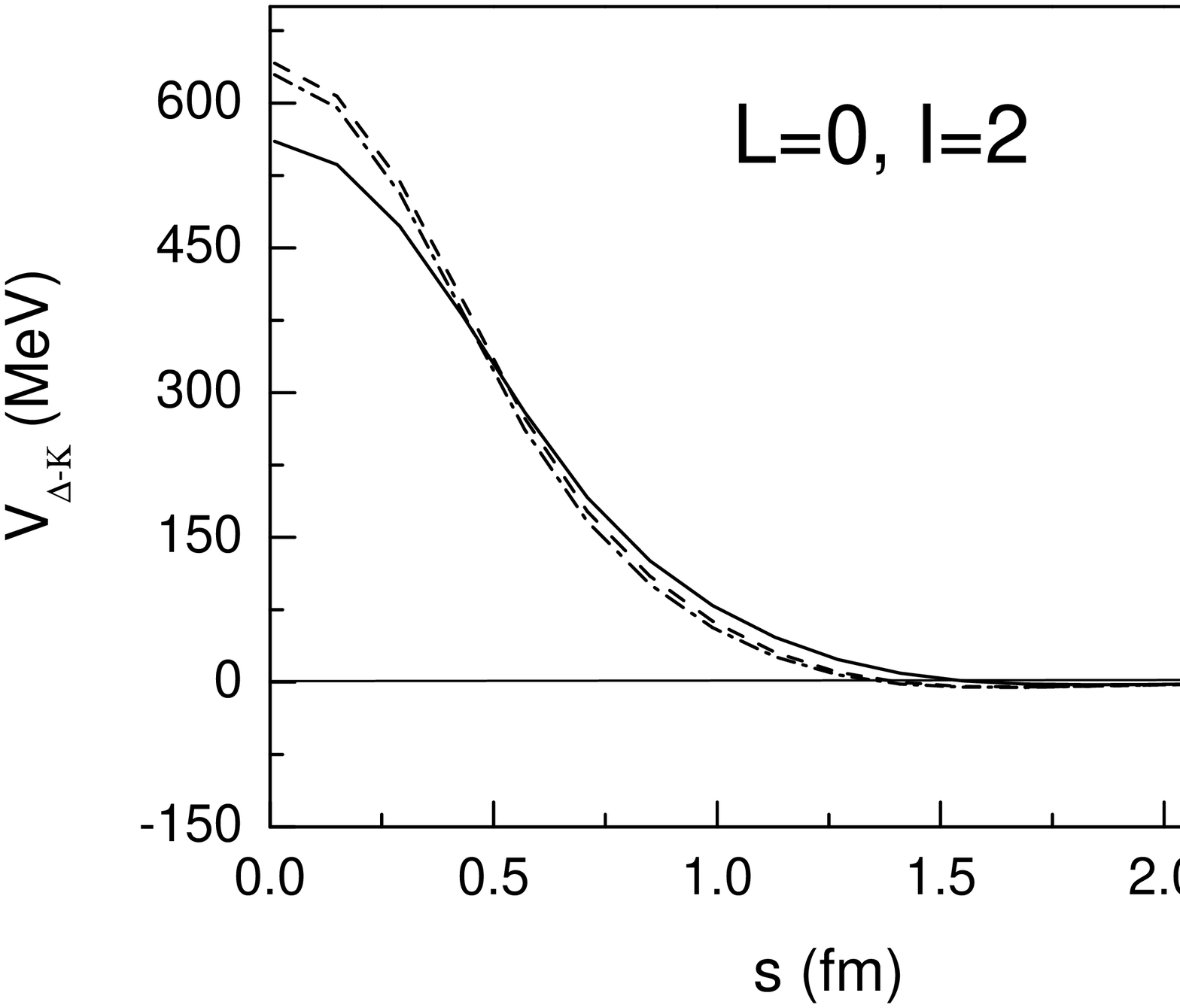,width=7.5cm} \vglue -2.3cm \caption{\small
The GCM matrix elements of the Hamiltonian. The solid curves
represent the results obtained in the chiral SU(3) quark model.
The dashed and dash-dotted curves show the results from the
extended chiral SU(3) quark model by taking $f_{chv}/g_{chv}$ as
$0$ and $2/3$, respectively.}
\end{figure}

In this work, we further dynamically study the $\Delta K$ state in
the extended chiral SU(3) quark model, where the vector-meson
exchanges play an important role in the short-range interaction.
Figure 1 shows the diagonal matrix elements of the Hamiltonian in
the generator coordinate method (GCM) \cite{kwi77} calculation,
which can describe the interaction between two clusters $\Delta$
and $K$ qualitatively. In Fig. 1, $s$ denotes the generator
coordinate and $V_{\Delta\mbox{-}K}$ is the effective potential
between the two clusters. From Fig. 1, one sees that the $\Delta
K$ state with isospin $I=1$ has an attractive interaction. Such an
attraction can consequently make for a $\Delta K$ bound state, and
the binding energy is tabulated in Table II. As can be seen in
Fig. 1, the $\Delta K$ interaction for the isospin $I=1$ channel
is more attractive in the extended chiral SU(3) quark model than
that in the original chiral SU(3) quark model, and thus the cases
II and III give much bigger binding energy than that of case I. In
the original chiral SU(3) quark model, the $\Delta K$ attraction
comes from the $\sigma$ exchange and the color-magnetic force of
OGE. In the extended chiral SU(3) quark model, the OGE is nearly
replaced by the vector-meson exchanges and the attraction
dominantly comes from the $\sigma$ and $\rho$ exchanges.

{\small
\begin{table}[htb]
\caption{Binding energy of $\Delta K$.}
\begin{center}
\begin{tabular*}{80mm}{@{\extracolsep\fill}ccc}
\hline\hline
 Model    & $B_{\Delta K}$ (MeV)  & Attraction \\
\hline
 I   &   3   &  OGE+$\sigma$ \\
 II  &  20   &  $\sigma$ + $\rho$   \\
 III &  15   &  $\sigma$ + $\rho$ \\
\hline\hline
\end{tabular*}
\end{center}
\end{table}}

Since the kaon meson is spin zero, the tensor force that plays an
important role in reproducing the binding energy of the deuteron
\cite{lrdai03} now nearly vanishes in the $\Delta K$ system. To
examine whether ($\Delta K$)$_{LSJ=0\frac{3}{2}\frac{3}{2}}$ is a
possible resonance or bound state, the channel coupling between
($\Delta K$)$_{LSJ=0\frac{3}{2}\frac{3}{2}}$ and
($NK^*$)$_{LSJ=0\frac{3}{2}\frac{3}{2}}$ will be considered in
future work.

\begin{figure}[htb]
\epsfig{file=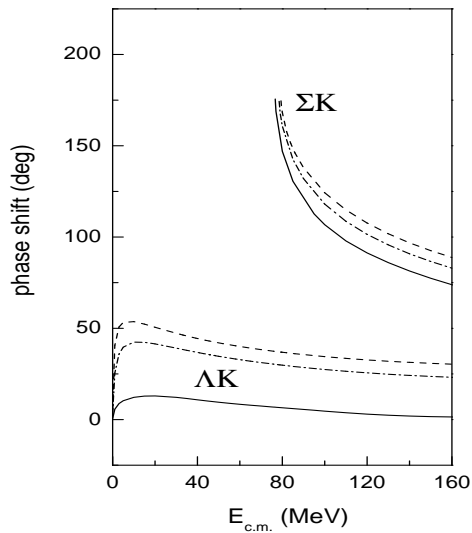,width=6.5cm,height=7.5cm} \caption{\small
The $S$-wave $\Lambda K$ and $\Sigma K$ phase shifts in the
one-channel calculation. The notation is the same as in Fig. 1.}
\end{figure}

The highlight that attracts our attention to the study of the
$\Lambda K$ system is the nucleon resonance $S_{11}(1535)$, of
which the traditional picture is that of an excited three quark
state, with one of the three quarks orbiting in an $l=1$ state
around the other two \cite{isgur78,glozman96}. In contrast from
the description in the constituent quark model (CQM), on the
hadron level the $S_{11}(1535)$ is argued to be a quasibound
$\Lambda K$-$\Sigma K$ state \cite{kaiser95,inoue02}.
Nevertheless, in Ref. \cite{schutz98}, the authors conclude that
the $S_{11}(1535)$ is not only generated by coupling to higher
baryon-meson channels but appears to require a genuine three-quark
component. So up to now the physical nature of the $S_{11}(1535)$
--- whether it is an excited three quark state or a quasi-bound
baryon-meson $S$-wave resonance or a mixing of these two
possibilities --- is still a stimulating problem. A dynamical
study on a quark level of the $\Lambda K$ and $\Sigma K$
interactions will undoubtedly make for a better understanding of
the $S_{11}(1535)$ and $S_{11}(1650)$.

{\small
\begin{table}[htb]
\caption{Binding energy of $\Sigma K$.}
\begin{center}
\begin{tabular*}{80mm}{@{\extracolsep\fill}ccc}
\hline\hline
 Model    & $B_{\Sigma K}$ (MeV)  & Attraction \\
\hline
I   &   18    &  OGE+$\sigma$ \\
II  &   44    &  $\sigma$ + $\rho$ + $\phi$  \\
III &   33    &  $\sigma$ + $\rho$ + $\phi$ \\
\hline\hline
\end{tabular*}
\end{center}
\end{table}}

In Ref. \cite{fhuang05lksk}, we study the $\Lambda K$ and $\Sigma
K$ states in the chiral SU(3) quark model and find a strong
attraction between $\Sigma$ and $K$, which consequently results in
a $\Sigma K$ quasi-bound state with about $17$ MeV binding energy.
When the channel coupling of $\Lambda K$ and $\Sigma K$ is
considered, a sharp resonance appears with spin-parity
$J^P=1/2^-$. Further analysis shows that the OGE plays an
important role in the $\Lambda K$ and $\Sigma K$ systems.

\begin{figure}[htb]
\epsfig{file=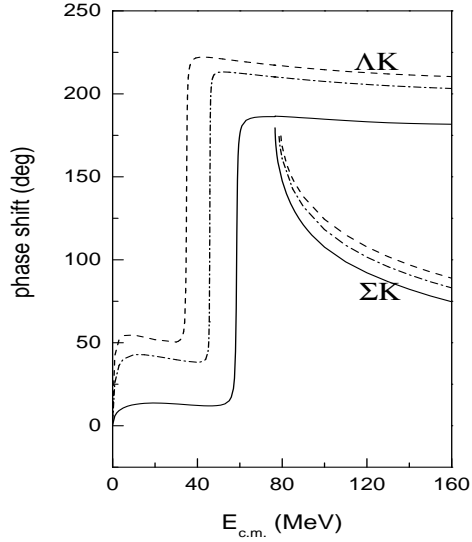,width=6.5cm,height=7.5cm}
\caption{\small The $S$-wave $\Lambda K$ and $\Sigma K$ phase
shifts in the coupled-channel calculation. The notation is the
same as in Fig. 1.}
\end{figure}

In this work, we further study the $\Lambda K$ and $\Sigma K$
systems in the extended chiral SU(3) quark model where the OGE is
nearly reduced. Figure 2 shows the $\Lambda K$ and $\Sigma K$
scattering phase shifts in the one-channel calculation. The phase
shifts denote that the $\Sigma K$ state has a strong attractive
interaction, which is consistent with the chiral Lagrangian
calculation on the hadron level \cite{kaiser95}. Such an
attraction can result in a $\Sigma K$ bound state, and the binding
energy is tabulated in Table III. Similar to the $\Delta K$
system, the interaction of $\Sigma K$ is more attractive in the
extended chiral SU(3) quark model than that in the original chiral
SU(3) quark model, and thus model II and model III give much
bigger binding energy than model I. In the original chiral SU(3)
quark model, the $\Sigma K$ attractive interaction comes from the
$\sigma$ exchange and the color-magnetic force of OGE. In the
extended chiral SU(3) quark model, the OGE is nearly replaced by
the vector-meson exchanges and the attraction dominantly comes
from the $\sigma$, $\rho$, and $\phi$ exchanges.

{\small
\begin{table}[htb]
\caption{Mass and width of the $\Lambda K$-$\Sigma K$ resonance.}
\begin{center}
\begin{tabular*}{80mm}{@{\extracolsep\fill}ccc}
\hline\hline
 Model    & Mass (MeV)  & $\Gamma$ (MeV) \\
\hline
 I   &  1670   &  $\approx 5$ \\
 II  &  1646   &  $\approx 4$  \\
 III &  1655   &  $\approx 4$ \\
\hline\hline
\end{tabular*}
\end{center}
\end{table}}

We also consider the channel coupling of $\Lambda K$ and $\Sigma
K$, the phase shifts of which are shown in Fig. 3. One sees that
there is a sharp resonance between the thresholds of $\Lambda K$
and $\Sigma K$. The narrow gap of the $\Lambda K$ and $\Sigma K$
thresholds, the strong attraction between $\Sigma$ and $K$, and
the sizeable off-diagonal matrix elements between $\Lambda K$ and
$\Sigma K$ are responsible for the appearance of this resonance.
The spin-parity of this resonance is $J^P=1/2^-$, and its mass and
width are tabulated in Table IV. The results from the extended
chiral SU(3) quark model are quite similar to those from the
original chiral SU(3) quark model, because $\rho$ and $\phi$
exchanges make contributions similar to OGE in this case. From the
mass point of view and considering that the branching ratio of
$S_{11}(1650)$ to $\Lambda K$ is $3-11\%$ (With a partial width of
about $4.5-16.5$ MeV), the resonance we obtained seems to be an
$S_{11}(1650)$, although the calculated width is a little bit
small. To draw a final conclusion regarding what the resonance we
obtained is and its exact theoretical mass and width, the effects
of the $s$-channel $q\bar q$ annihilation interaction as well as
the coupling to the $N\pi$, $N\eta$, $N\pi\pi$, and even to the
genuine $3q$ component will be considered in future work.

In summary, we dynamically study the $\Delta K$, $\Lambda K$, and
$\Sigma K$ states in the extended chiral SU(3) quark model, where
the coupling between the quark and vector chiral fields are
considered and thus the OGE is nearly reduced. Although the
mechanisms of the quark-quark short-range interactions are quite
different in the original chiral SU(3) quark model and the
extended chiral SU(3) quark model, the theoretical results from
these two models are very similar in these cases. They both show
that the interactions of $\Delta K$ with isospin $I=1$ and $\Sigma
K$ with isospin $I=1/2$ are attractive, which can consequently
lead to $\Delta K$ and $\Sigma K$ quasibound states. When the
channel coupling of $\Lambda K$ and $\Sigma K$ is considered, our
calculated phase shifts show a sharp resonance between the
thresholds of these two channels with spin-parity $J^P=1/2^-$. Its
exact theoretical mass and width await future work where more
channel couplings and the decay properties will be studied.

\vspace{0.5cm}

This work was supported in part by the National Natural Science
Foundation of China, Grant No. 10475087.

\end{document}